\documentclass[a4paper,twoside,twocolumn,english,aps,superscriptaddress,showpacs]{revtex4}
\usepackage{times}
\usepackage[T1]{fontenc}
\usepackage[latin1]{inputenc}
\usepackage{subfigure}
\usepackage{amsmath}
\usepackage{graphicx}
\usepackage{amssymb}

\makeatletter

\newcommand{\lyxdot}{.}

\usepackage{babel}
\makeatother
\begin{document}

\title{Excitations of attractive 1-D bosons: Binding vs. fermionization}

\author{Emmerich Tempfli}

\affiliation{Theoretische Chemie, Universit\"{a}t Heidelberg, Im Neuenheimer
Feld 229, 69120 Heidelberg, Germany}

\author{Sascha Zöllner}

\email{sascha.zoellner@pci.uni-heidelberg.de}

\affiliation{Theoretische Chemie, Universit\"{a}t Heidelberg, Im Neuenheimer
Feld 229, 69120 Heidelberg, Germany}

\author{Peter Schmelcher}

\email{peter.schmelcher@pci.uni-heidelberg.de}

\affiliation{Theoretische Chemie, Universit\"{a}t Heidelberg, Im Neuenheimer
Feld 229, 69120 Heidelberg, Germany}

\affiliation{Physikalisches Institut, Universit\"{a}t Heidelberg, Philosophenweg
12, 69120 Heidelberg, Germany}


\date{June 5, 2008}

\begin{abstract}
The stationary states of few bosons in a one-dimensional harmonic
trap are investigated throughout the crossover from weak to strongly
attractive interactions. For sufficient attraction, three different
classes of states emerge: (i) $N$-body bound states, (ii) bound states
of smaller fragments, and (iii) gas-like states that fermionize, that
is, map to ideal fermions in the limit of infinite attraction. The
two-body correlations and momentum spectra characteristic of the three
classes are discussed, and the results are illustrated using the soluble
two-particle model.
\end{abstract}
\maketitle

\section{Introduction}

In recent years, ultracold atoms have become a flexible tool for the
simulation of fundamental quantum systems \cite{pethick,pitaevskii,bloch07}.
Their versatility derives mainly from the fact that both their external
forces and atomic interactions can be designed to a great extent.
One striking example is the possibility to confine the atoms' motion
to lower dimensions, such as in the one-dimensional (1D) Bose gas
\cite{bloch07}. Since, pictorially speaking, particles moving on
a line cannot move around each other, they are in a sense more strongly
correlated than their higher-dimensional counterparts. Moreover, their
effective interaction strength can be tuned freely so as to enter
the interesting regime of strong interactions, either via Feshbach
resonances of the 3D scattering length \cite{koehler06} or through
confinement-induced resonances of the effective 1D coupling \cite{Olshanii1998a}.

The case of repulsive interactions has long received considerable
attention, mostly for the striking feature that, in the hard-core
limit of infinite repulsion between the bosons, the system maps to
an ideal Fermi gas \cite{girardeau60}. In this \emph{fermionization}
limit, the bosons become impenetrable, which has a similar effect
as Pauli's exclusion principle for identical fermions.  The seminal
exact solutions derived for special systems at arbitrary interaction
strength---such as the \emph{homogeneous} Bose gas on a ring in the
thermodynamic limit \cite{lieb63a,lieb63b} as well as for definite
particle numbers \cite{sakmann05}, and for the \emph{inhomogeneous}
gas in a hard-wall trap \cite{hao06}---have recovered this borderline
case in the limit of infinite coupling. The thermodynamic nature of
the fermionization crossover from weak to strong repulsion has first
been explored and contrasted with the complementary Thomas-Fermi regime
\cite{petrov00,dunjko01}, and its microscopic mechanism has been
unraveled by recent numerically exact studies \cite{deuretzbacher06,zoellner06a,zoellner06b,schmidt07}.
Moreover, its experimental demonstration has sparked renewed interest
in 1D bosons \cite{kinoshita04,paredes04}.

By contrast, the understanding of the attractive case is more patchy.
In the homogeneous system, the ground state forms an $N$-body bound
state \cite{mcguire64}. For sufficient attraction, it becomes ever
more localized and is unstable in the thermodynamic limit. For finite
systems, though, the ground state remains stable for arbitrary finite
attraction, as is demonstrated by the exact solution via Bethe's ansatz
for a ring \cite{sakmann05} and a hard-wall trap \cite{hao06}. Much
less is known about excited states. In the homogeneous system again,
Monte Carlo simulations have indicated the existence of a highly excited
\emph{gas-like} state for ultrastrong attraction, which can be seen
as the counterpart of the fermionized ground state for \emph{repulsive}
interactions \cite{astrakharchik05,astrakharchik04a}. The evidence
for this \emph{super-Tonks} gas has been supported by a Bethe-ansatz
solution \cite{batchelor05}. Still, an intuitive understanding of
these states from a microscopic perspective and how they come about
in the crossover from weak interactions is still missing. The complementary
crossover for the \emph{low-lying} excitation spectrum in turn has
been investigated recently for the homogeneous system \cite{sykes07};
however, it does not include the gas-like super-Tonks gas.

In this article, we study the entire crossover from the non-interacting
to the strongly attractive limit for few bosons in a harmonic trap.
This is done via the numerically exact multi-configuration time-dependent
Hartree method introduced in Sec.~\ref{sec:method}. Section~\ref{sec:theory}
presents the general Bose-Fermi map valid for the gas-like super-Tonks
states, and illustrates its meaning on the simple model of two bosons.
The numerical investigation of the stationary states in Sec.~\ref{sec:mol-vs-fermi}
reveals three distinct classes for strong enough attraction: $N$-body
bound states (Sec.~\ref{sub:Trimer}), states involving smaller fragments
(Sec.~\ref{sub:Hybrid}), and finally gas-like states that fermionize
(Sec.~\ref{sub:Fermionizing-states}).

\section{Model and computational method\label{sec:method}}

\paragraph*{Model}

We consider $N$ trapped bosons described by the Hamiltonian \[
H=\sum_{i=1}^{N}\left[\frac{1}{2}p_{i}^{2}+U(x_{i})\right]+g\sum_{i<j}\delta_{\sigma}(x_{i}-x_{j}).\]
We will focus on the case of harmonic confinement, $U(x)=\frac{1}{2}x^{2}$
(where harmonic-oscillator units are employed throughout.) The effective
interaction resembles a 1D contact potential, but is mollified with
a Gaussian $\delta_{\sigma}(x)\equiv e^{-x^{2}/2\sigma^{2}}/\sqrt{2\pi}\sigma$
(of width $\sigma=0.05$) for numerical reasons (cf. \cite{zoellner06a}
for details.) We concentrate on attractive forces $g\in(-\infty,0]$,
which can be achieved experimentally by either having negative scattering
lengths or by reducing the transverse confinement length $a_{\perp}\equiv\sqrt{\hbar/M\omega_{\perp}}$
sufficiently \cite{Olshanii1998a}.

\paragraph*{Computational method }

Our approach relies on the numerically exact multi-configuration time-dependent
Hartree method \cite{mey90:73,bec00:1,mctdh:package}, a quantum-dynamics
approach which has been applied successfully to systems of few identical
bosons \cite{zoellner06a,zoellner06b,zoellner07a,zoellner07c,zoellner07b}
as well as to Bose-Bose mixtures \cite{zoellner08b}. Its principal
idea is to solve the time-dependent Schrödinger equation $\begin{array}{c}
i\dot{\Psi}(t)=H\Psi(t)\end{array}$ as an initial-value problem by expanding the solution in terms of
direct (or Hartree) products $\Phi_{J}\equiv\varphi_{j_{1}}\otimes\cdots\otimes\varphi_{j_{N}}$:\begin{equation}
\Psi(t)=\sum_{J}A_{J}(t)\Phi_{J}(t).\label{eq:mctdh-ansatz}\end{equation}
The unknown single-particle functions $\varphi_{j}$ ($j=1,\dots,n$)
are in turn represented in a fixed \emph{}basis of, in our case, harmonic-oscillator
orbitals. The permutation symmetry of $\Psi$ is ensured by the correct
symmetrization of the expansion coefficients $A_{J}$. 

Note that, in the above expansion, not only the coefficients $A_{J}$
but also the single-particle functions $\varphi_{j}$ are time dependent.
Using the Dirac-Frenkel variational principle, one can derive equations
of motion for both $A_{J},\varphi_{j}$ \cite{bec00:1}. Integrating
this differential-equation system allows us to obtain the time evolution
of the system via (\ref{eq:mctdh-ansatz}). This has the advantage
that the basis set $\{\Phi_{J}(t)\}$ is variationally optimal at
each time $t$; thus it can be kept relatively small. 

Although designed for time-dependent studies, it is also possible
to apply this approach to stationary states. This is done via the
so-called \emph{relaxation} method \cite{kos86:223}. The key idea
is to propagate some wave function $\Psi(0)$ by the non-unitary $e^{-H\tau}$
(propagation in imaginary time.) As $\tau\to\infty$, this exponentially
damps out any contribution but that stemming from the true ground
state like $e^{-(E_{m}-E_{0})\tau}$. In practice, one relies on a
more sophisticated scheme termed \emph{improved relaxation} \cite{mey03:251,meyer06},
which is much more robust especially for excitations. Here $\langle\Psi|H|\Psi\rangle$
is minimized with respect to both the coefficients $A_{J}$ and the
orbitals $\varphi_{j}$. The effective eigenvalue problems thus obtained
are then solved iteratively by first solving for $A_{J}$ with \emph{fixed}
orbitals and then {}`optimizing' $\varphi_{j}$ by propagating them
in imaginary time over a short period. That cycle will then be repeated.

\section{Bose-Fermi map for attractive bosons \label{sec:theory}}

In this section, we state the general Bose-Fermi map and discuss its
application to infinitely \emph{attractive} interactions. (Without
loss of generality, we focus on the time-independent formulation.)
Its intuitive meaning will be illustrated on the special example of
two harmonically trapped bosons.

\subsection{General map}

The Schrödinger equation of $N$ bosons with point interactions, $(E-H)\Psi=0$,
is equivalent to a noninteracting system $(E-H_{0})\Psi=0$ with boundary
conditions \begin{equation}
2\left.\partial_{r}\Psi\right|_{r=0^{+}}=\left.g\Psi\right|_{r=0},\label{eq:pp-BC}\end{equation}
 where $r\equiv x_{i}-x_{j}$ for fixed $i\neq j$. For infinitely
repulsive interactions, $g\to\infty$, the constraint that \[
\left.\frac{\partial_{r}\Psi}{\Psi}\right|_{r=0^{+}}\to+\infty\]
 leads to the well-known hard-core boundary conditions \begin{equation}
\left.\Psi\right|_{x_{i}=x_{j}}=0,\qquad i\neq j.\label{eq:hc-BC}\end{equation}
Since, otherwise, $\Psi$ fulfills the noninteracting Schrödinger
equation, it is intelligible that it can be mapped to a noninteracting
state $\Psi_{-}$ of identical \emph{fermions}, which automatically
satisfies the hard-core boundary condition (\ref{eq:hc-BC}) by Pauli's
exclusion principle \cite{girardeau60}:\[
\Psi=A\Psi_{-},\quad A(X)\equiv\prod_{i<j}\mathrm{sgn}(x_{i}-x_{j}).\]
The Bose-Fermi map $A$ serves only to restore bosonic permutation
symmetry. Note that, since $A^{2}=1$, all local quantities derived
from $\rho_{N}=\left|\Psi\right|^{2}$ will coincide with those computed
from the fermion state. In this sense, the case of infinite repulsion
is commonly referred to as \emph{fermionization} limit ({}``Tonks
gas'').

By contrast, the constraint for infinite attraction \[
\left.\frac{\partial_{r}\Psi}{\Psi}\right|_{r=0^{+}}\to-\infty\]
may be satisfied by (a) $\left.\partial_{r}\Psi\right|_{0^{+}}\to-\infty$
(provided $\left.\Psi\right|_{0}>0$ diverges more slowly than the
local derivative) or (b) $\left.\Psi\right|_{0}\to0^{-}$ (assuming
$\left.\partial_{r}\Psi\right|_{0^{+}}>0$) . If case (b) applies
\emph{}to \emph{all} $i\neq j$, then we recover the hard-core gas
above. Consequently, the Bose-Fermi mapping $\Psi=A\Psi_{-}$ then
holds for $g\to-\infty$ as well. In particular, the energetically
lowest such state exactly equals the \emph{fermionized} repulsive
ground state -- the Tonks gas. However, for strong attraction, this
eigenstate will be highly excited, whereas the ground state will be
strongly bound. For finite $g>-\infty$, this may be identified with
the \emph{super-Tonks} state in Ref.~\cite{astrakharchik05}.

A few comments are in order: 

(i) By construction, this holds for any external potential $U$, just
like the standard Bose-Fermi map. In particular, the analytic solution
for $N$ bosons in a harmonic trap carries right over \cite{girardeau01}\begin{equation}
\Psi\left(X\right)\propto e^{-|X|^{2}/2}\negthickspace\prod_{1\le i<j\le N}\negthickspace|x_{i}-x_{j}|,\label{eq:TG-HO}\end{equation}
and likewise for the homogeneous system \cite{girardeau60}. 

(ii) By the same logic as above, this extends to binary mixtures of
bosons (for the repulsive case, cf. \cite{girardeau07}). Likewise,
the generalized Bose-Fermi map for spinor bosons \cite{deuretzbacher08}
ought to apply also to the limit of infinite attraction.

(iii) The map also holds in the presence of additional long-range
interactions, such as in dipolar gases \cite{astrakharchik08}.

\paragraph*{}

\subsection{Illustration \label{sub:Illustration}}

To visualize the above argument, let us resort to the simple model
of two bosons in a harmonic-oscillator (HO) potential, $U(x)=\frac{1}{2}x^{2}$.
Here the center of mass (CM) $R=\frac{1}{N}\sum_{i}x_{i}$ and the
relative coordinate $r=x_{1}-x_{2}$ separate,\[
H=h_{\mathrm{CM}}+H_{\mathrm{rel}}\equiv\left[\frac{p_{R}^{2}}{2N}+\frac{1}{2}NR^{2}\right]+\left[p_{r}^{2}+\frac{1}{4}r^{2}+g\delta(r)\right]\!.\]
One can therefore write the wave function and its energy as\[
\Psi=\phi_{\mathcal{N}}\otimes\psi;\quad E=(\mathcal{N}+{\scriptstyle \frac{1}{2}})+\epsilon_{\mathrm{}},\]
where $\phi_{\mathcal{N}}$ is the HO orbital with quantum number
$\mathcal{N}=0,1,\dots$ . The relative Hamiltonian may be viewed
as a harmonic potential split into halves in the center, i.e., at
the point of collision $r=0$. There the delta function imposes the
boundary condition (\ref{eq:pp-BC}), which amounts to a cusp for
$g<0$. The problem can be solved analytically in terms of parabolic
cylinder functions $U(a,b)$ \cite{Busch98,cirone01} \begin{equation}
\psi_{\epsilon}(r)=cU(-\epsilon,r),\label{eq:pp-psi}\end{equation}
where $\epsilon(g)\equiv\nu(g)+\frac{1}{2}$ is determined through
the transcendental equation \begin{equation}
\nu(g)\in f_{g}^{-1}(0):\quad f_{g}(\nu):=2^{3/2}\frac{\Gamma\left(\frac{1-\nu}{2}\right)}{\Gamma\left(-\frac{\nu}{2}\right)}+g.\label{eq:pp-energy}\end{equation}

The solution for attractive interactions is plotted in Fig.~\ref{fig:2p_HO-states}:
\begin{figure}
\begin{centering}\subfigure[]{\includegraphics[width=0.5\columnwidth]{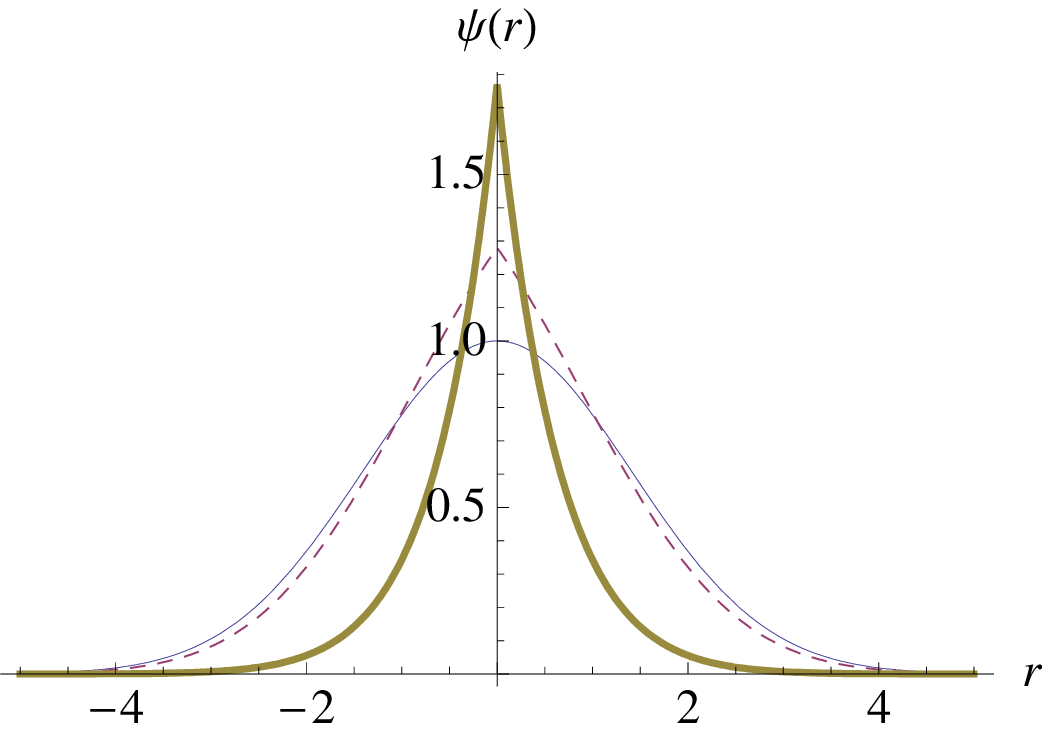}}\subfigure[]{\includegraphics[width=0.5\columnwidth]{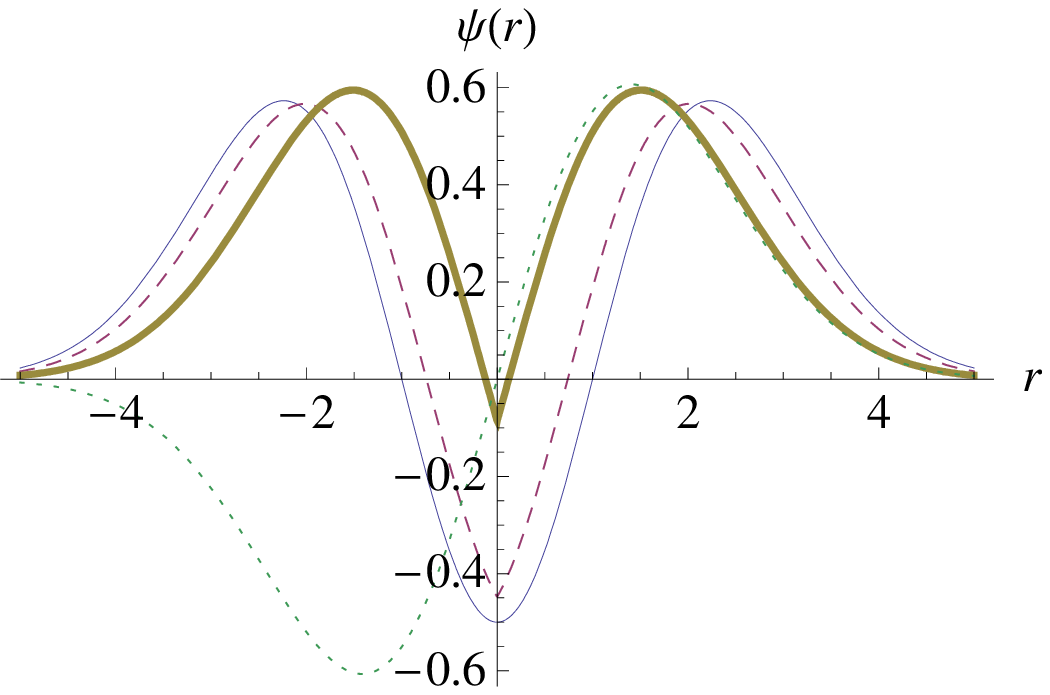}}\par\end{centering}

\caption{(color online) Relative wave function $\psi_{\nu(g)}(r)$ of two
bosons in a harmonic trap (in arbitrary units).\protect \\
(a) \emph{Ground state for attractive interactions}: The Gaussian
ground state ($\nu(g=0)=0$, thin-blue line) becomes peaked for $g=-0.6$
(dashed-magenta). This tends to an exponential peak as $g\to-\infty$
(cf. $g=-3.2$, thick-ocher line).\protect \\
 (b) \emph{Excited states for attractive interactions}: The first
bosonic excitation ($\nu=2$, thin-blue line) picks up a cusp at zero
for $g=-1.5$ (dashed-magenta). As $g\to-\infty$ (cf. $g=-16$, thick-ocher),
this cusp becomes sharper but, at the same time, is damped out more
and more until the wave function's modulus equals the fermionic state
$u_{1}(r)\propto r\, e^{-r^{2}/4}$ (dotted-green). \label{fig:2p_HO-states}}
\end{figure}

\begin{itemize}
\item First imagine we start from the noninteracting ground state $\nu(0)=0$
(Fig.~\ref{fig:2p_HO-states}a). For $g<0$, the Gaussian will pick
up a cusp at $r=0$, which becomes ever sharper for increasing $\left|g\right|$,
i.e., $\psi'(0^{+})\to-\infty$. In the limit $g\to-\infty$, when
the support of $\psi$ becomes much smaller than the oscillator length,
this is described simply by the bound state of the delta potential,
$\psi(r)\sim\frac{1}{\sqrt{a}}e^{-\left|r\right|/a}$ with $a\equiv-2/g$
the 1D scattering length. Clearly $\left|\psi(r)\right|^{2}\stackrel{g\to-\infty}{\to}$$\delta(r)$. 
\item An eigenstate with $\nu(0)=2n\neq0$ will also form a cusp at $r=0$
(Fig.~\ref{fig:2p_HO-states}b). In contrast to before, however,
this peak will become ever smaller for $g\to-\infty$, until it reaches
down to $\psi(0)=0$ for infinite attraction. In this limit, the wave
function's modulus equals that of the next lower fermionic state,
$\psi(r)\to\mathrm{sgn}(r)u_{2n-1}(r)$, where $u_{a}(r)\propto H_{a}\left(\frac{r}{\sqrt{2}}\right)\, e^{-r^{2}/4}$.
Concomitantly, the energy will be lowered to $\lim_{g\to-\infty}\nu(g)=2n-1$,
which matches exactly the fermionized level $\lim_{g\to+\infty}\bar{\nu}(g)$
starting from $\bar{\nu}(0)=2(n-1)$. 
\end{itemize}

\section{Molecule formation vs. fermionization in a harmonic trap \label{sec:mol-vs-fermi}}

After having presented the general mapping valid for infinite attraction,
let us now investigate the \emph{crossover} to that borderline case,
starting from the noninteracting states. For concreteness, we shall
focus on the case of $N=3$ bosons in a harmonic trap. 

\begin{figure}
\begin{centering}\includegraphics[width=1\columnwidth,keepaspectratio]{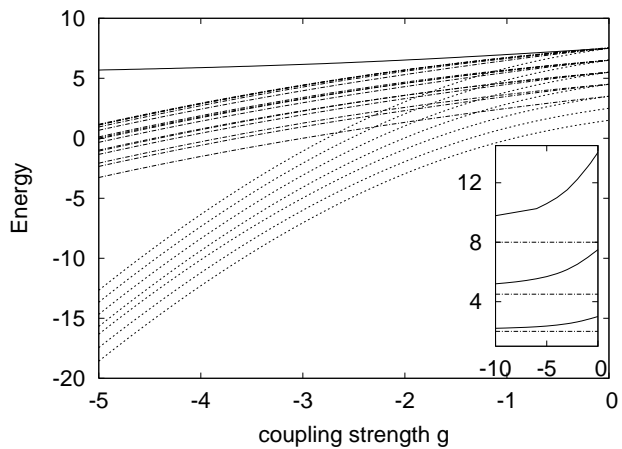}\par\end{centering}

\caption{Energy spectrum $\{ E_{m}(g)\}$ of $N=3$ bosons in a harmonic trap.
Different line styles correspond to states which connect to a trimer
(- - -), a dimer plus a single boson ($-\cdot-$), and a gas-like
fermionized state (---), respectively. \emph{Inset}: Energy $E(g)$
for the lowest fermionized states of $N=2-4$ atoms (\emph{bottom
to top}). \label{cap:energy}}
\end{figure}

Figure~\ref{cap:energy} explores the evolution of the low-lying
few-body spectrum $\{ E_{m}(g)\}$ as we vary $g\le0$. In the absence
of interactions, the spectrum exhibits an even spacing, which comes
about by distributing all $N$ particles in number states $|\boldsymbol{N}\rangle=|N_{0},N_{1},\dots\rangle$
over the lowest HO levels $\epsilon_{a}=a+\frac{1}{2}$. Clearly,
Fig.~\ref{cap:energy} reveals that, when attractive interactions
$g<0$ are switched on, the depicted spectrum falls apart into three
(for $N=3$) qualitatively different subclasses. In anticipation of
our analysis below, we identify the asymptotically lowest set of levels
as $N$-body bound states (or trimers), the cluster on top of these
as hybrid states (dimers plus one free boson), and the highest level
as gas-like state which undergoes fermionization. Note that, by CM
separation, for each state there is a countable set of copies with
different CM energies $\mathcal{N}+\frac{1}{2}$, which are shifted
with respect to one another.

\subsection{Trimer states \label{sub:Trimer}}

\begin{figure*}[!t]
\subfigure[]{\includegraphics[width=0.33\textwidth,keepaspectratio]{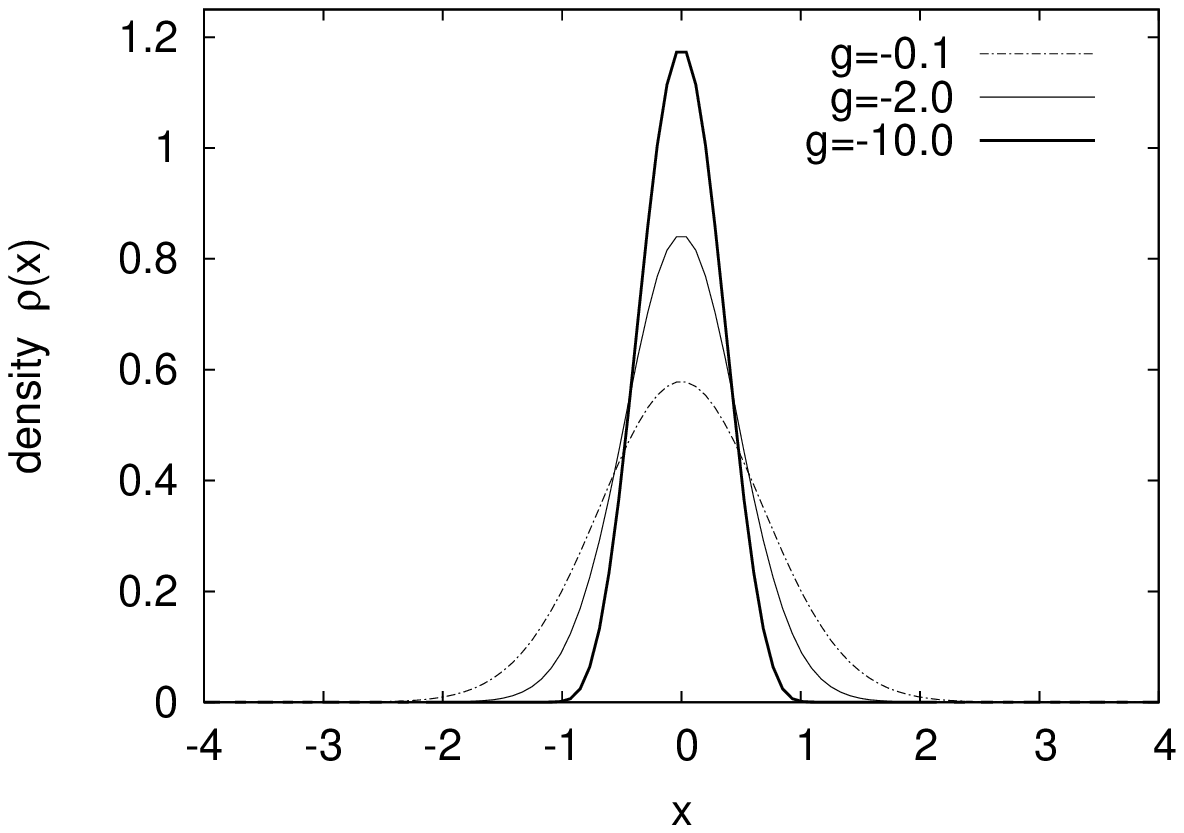}}\subfigure[]{\includegraphics[width=0.33\textwidth,keepaspectratio]{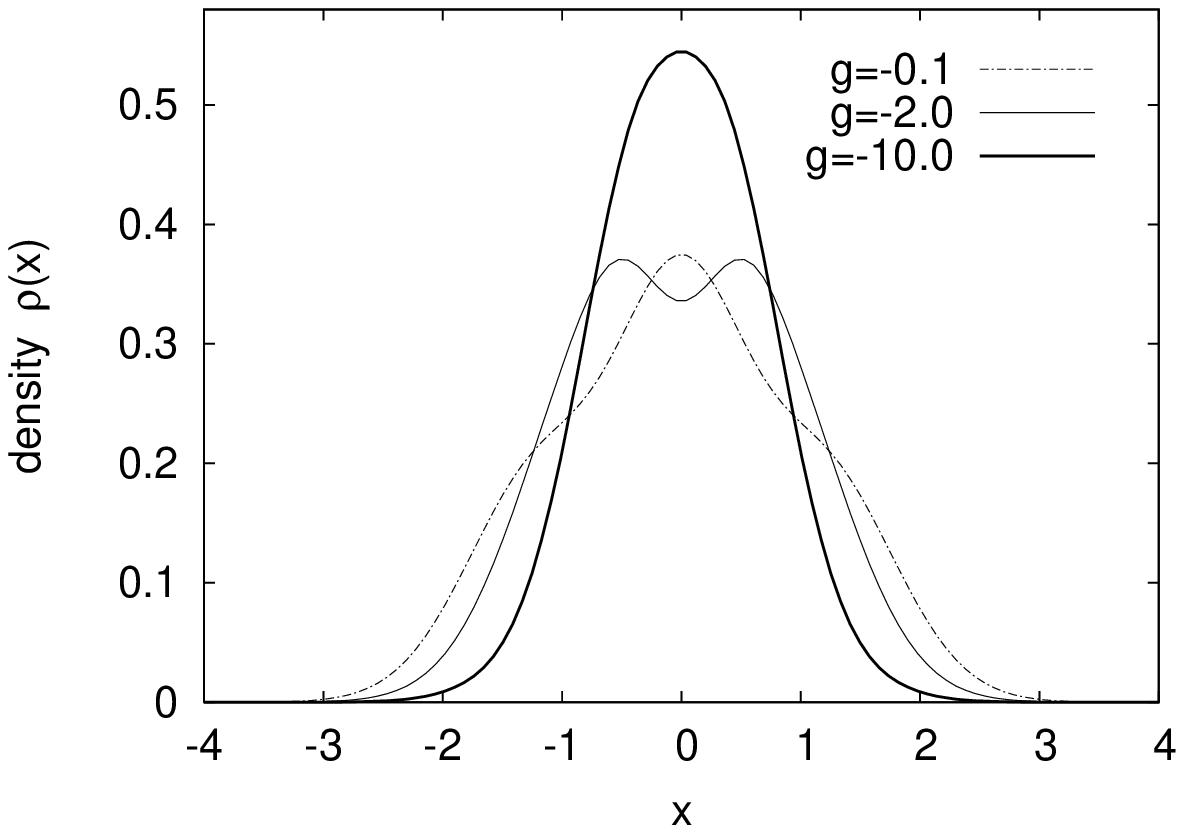}}\subfigure[]{\includegraphics[width=0.33\textwidth,keepaspectratio]{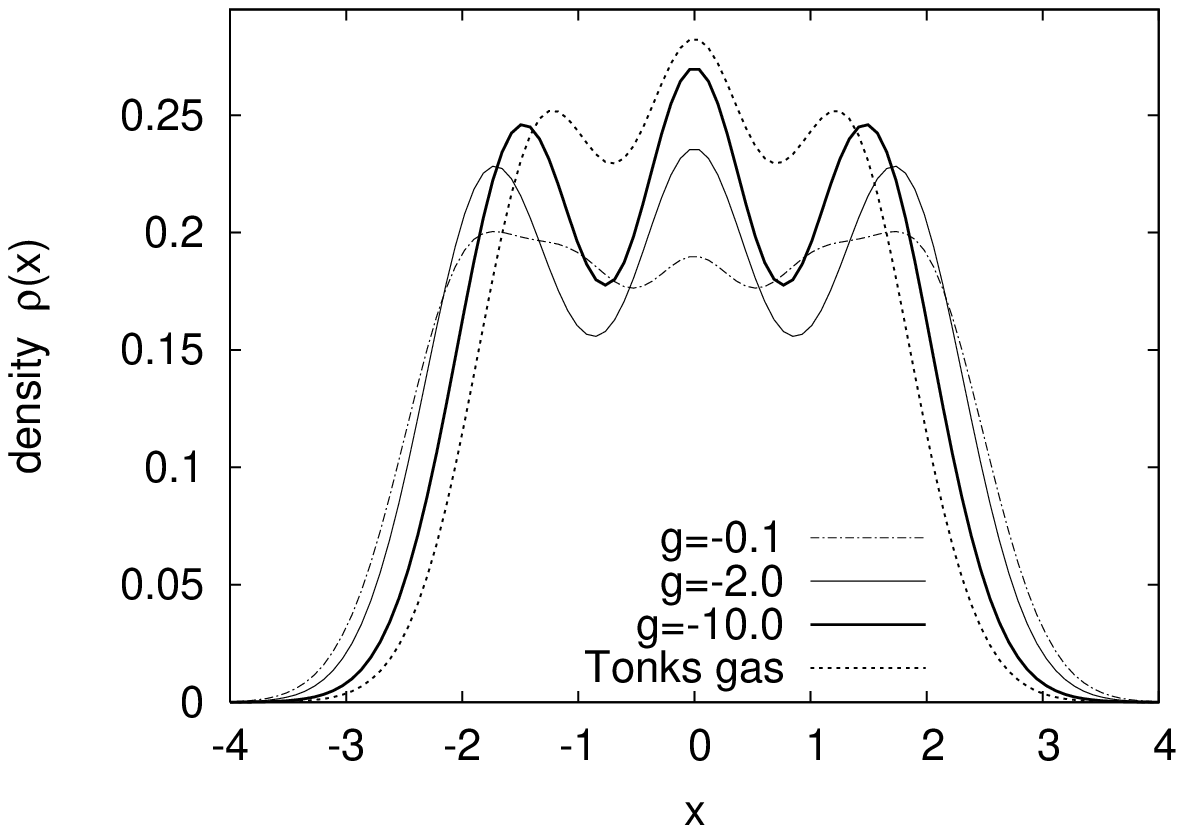}}

\caption{One-body density $\rho(x)$ for the ground state (a), some hybrid
state (b), and the lowest fermionizing state \textbf{(}c) of $N=3$
bosons, plotted for different interaction strengths $g$. \label{cap:density}}
\end{figure*}

The ground state, as well as its copies with higher CM excitations,
can be thought of as an $N$-body bound state, which would be the
straightforward generalization of the two-body bound state discussed
in Sec.~\ref{sub:Illustration}. In fact, this class of states is
well known from the homogeneous system \cite{mcguire64}, where an
analytic solution is available via Bethe's ansatz \cite{sakmann05,sykes07}\begin{equation}
E(g)\propto-cg^{2}N(N^{2}-1);\quad\Psi(X)\propto e^{-\sum_{i<j}|x_{i}-x_{j}|/a},\label{eq:N-bound}\end{equation}
where $a\equiv-2/g$ is the 1D scattering length. We will now argue
that, for sufficient attraction, this wave function also holds in
our case of harmonic confinement, up to a trivial CM factor $\phi_{\mathcal{N}}$.
To this end, let us proceed like in \cite{mcguire64} and transform
$X\equiv(x_{1},\dots,x_{N})^{\top}$ to Jacobian coordinates $Y\equiv(R,r_{1},,\dots,r_{N-1})^{\top}$,
here for simplicity specified for $N=3$: \[
Y=\mathcal{O}X,\;\mathcal{O}=\left(\begin{array}{ccc}
\frac{1}{\sqrt{3}} & \frac{1}{\sqrt{3}} & \frac{1}{\sqrt{3}}\\
\frac{1}{\sqrt{2}} & \frac{-1}{\sqrt{2}} & 0\\
\sqrt{\frac{2}{3}}\frac{1}{2} & \sqrt{\frac{2}{3}}\frac{1}{2} & -\sqrt{\frac{2}{3}}\end{array}\right).\]
Up to a factor, $R$ ($r_{1}$) coincide with the usual center of
mass (two-particle relative coordinate), while $r_{2}$ gives the
difference between particle $\#3$ and the center of mass of the cluster
$(1,2)$. By orthogonality of $\mathcal{O}$, the Hamiltonian transforms
to $H(Y)=h_{CM}+H_{\mathrm{rel}}$, with \begin{eqnarray}
H_{\mathrm{rel}} & = & \sum_{k=1}^{2}\left[\frac{1}{2}p_{r,k}^{2}+\frac{1}{2}r_{k}^{2}\right]+\label{eq:H_Jacobian}\\
 &  & g\left\{ \frac{1}{\sqrt{2}}\delta(r_{1})+\sqrt{2}\sum_{\pm}\delta(r_{1}\pm\sqrt{3}r_{2})\right\} .\nonumber \end{eqnarray}
If all $N$ particles cling together to form a tightly bound state,
then their distances will be small compared to the confinement scale,
$\left|r_{k}\right|\ll1$. In this limit, $\frac{1}{2}r_{k}^{2}$
may be safely neglected, so that the relative wave function asymptotically
maps to the homogeneous form (\ref{eq:N-bound}). Likewise, the energy
scales as $E(g)\sim\mathcal{N}+\frac{N}{2}-\alpha_{N}g^{2}$ ($\alpha_{N}>0$). 

\begin{figure}
\begin{centering}\includegraphics[width=0.9\columnwidth,keepaspectratio]{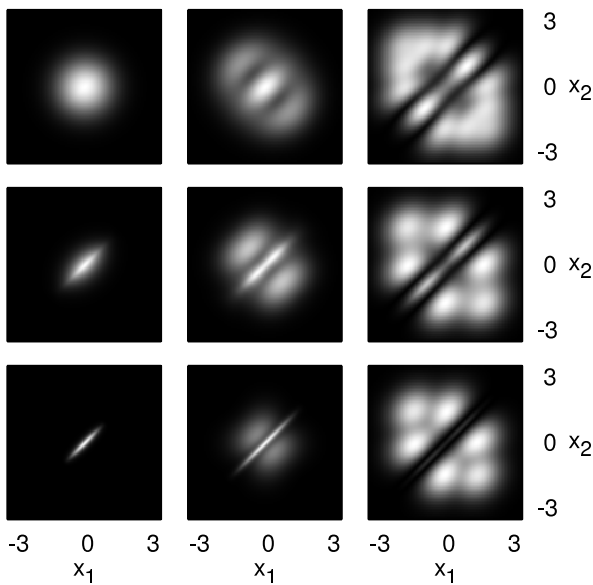}\par\end{centering}

\caption{Two-body correlation functions $\rho_{2}(x_{1},x_{2})$ for the ground
state, some hybrid state, and the lowest fermionizing state (\emph{from
left to right}), at coupling strengths $g=-0.1,\,-2$ and $-10$ (\emph{from
top to bottom}). \label{cap:2p-corr}}
\end{figure}

A look at the one-body density profiles $\rho(x)=\langle x|\rho_{1}|x\rangle$
shown in Fig.~\ref{cap:density}(a) confirms that this state becomes
more and more localized. In contrast to the translation-invariant
homogeneous case, this {}``soliton''-like state is localized even
in the one-body density, which represents an average over all measurements.
Moreover, in contrast to the hard-wall trap \cite{hao06}, there is
\emph{no} tipping point where the width $(\Delta x)^{2}=\langle x^{2}\rangle$
would become \emph{larger} again and return to its \emph{noninteracting}
value for $g\to-\infty$. In fact, here $\Delta x$ decreases monotonically
for $g\to-\infty$ and eventually saturates in the CM density \begin{equation}
\lim_{g\to-\infty}\rho(x)=\left|\phi_{\mathcal{N}}(x)\right|^{2},\label{eq:rho-bound}\end{equation}
whose length scale $a_{\mathrm{CM}}=1/\sqrt{N}$ is suppressed by
the total mass $NM$ (in units of the atomic mass $M\equiv1$). To
prove this, note that by CM separation, \[
\rho(x_{1})=\int\negmedspace dx_{2}\cdots dx_{N}\left|\phi_{\mathcal{N}}\left(R\right)\right|^{2}\negmedspace\rho_{\mathrm{rel}}(r\equiv\{ r_{k}\}).\]
For increasing attraction, $\lim_{g\to-\infty}\rho_{\mathrm{rel}}(r)=\delta(r)$,
in agreement with (\ref{eq:N-bound}). Fig.~\ref{cap:2p-corr} visualizes
that the two-body density $\rho_{2}(x_{1},x_{2})=\langle x_{1},x_{2}|\rho_{2}|x_{1},x_{2}\rangle$
is more and more concentrated on the diagonal $\{ x_{1}=x_{2}\}$.
Carrying out the trivial integrals proves Eq.~(\ref{eq:rho-bound}),
which reflects that all atoms clump together to point-like molecule
whose position coincides with the CM. (Of course, the validity of
our effective model requires the molecule to be still large compared
to the spatial extension of the atoms.)

That line of reasoning fails in the hard-wall trap, where $R$ and
$r$ couple strongly due to the anharmonicity of $U(x)$ \cite{matthies07};
in fact, for very tight binding of two particles, their common CM
would be permitted to spread out over the whole box, thus compensating
the stronger localization in the relative coordinate. A simple dimensional
argument to see this is that the length scale in a hard-wall trap
is simply the \emph{size} of the box, independent of the object's
mass -- in contrast to the harmonic oscillator.

\subsection{Hybrid states\label{sub:Hybrid}}

The behavior of the class of levels on top of the trimer levels ($N=3$,
Fig.~\ref{cap:energy}) is clearly more complex, which indicates
that different CM excitations $\mathcal{N}$ are involved. On the
one hand, the levels are significantly above those of the $N$-body
bound states; on the other hand, they do not saturate with increasing
attraction. This suggests to identify these three-boson states with
the formation of dimers plus one unbound atom. For general $N$, this
class of hybrid states involves different fragments---labeled $f=1,\dots,F$---of
$N_{f}$-body bound states, where $\sum_{f=1}^{F}N_{f}=N$ and $F=2,\dots,N-1$.

Figure~\ref{cap:density}(b) shows the density profile of the lowest
hybrid state, which has $\mathcal{N}=0$. A pronounced peak at $x=0$
builds up similar to the trimer case, but with a non-Gaussian structure
indicative of relative excitations. This becomes clearer from a look
at the two-body density in Fig.~\ref{cap:2p-corr}: For increasing
attraction, part of the bosons clump together, as is visible from
the {}``molecule'' peak at $\{ x_{1}=x_{2}\}$. However, unlike
before, a non-negligible part remains isolated at $\{ x_{1}=-x_{2}\}$.
To understand this pattern a little better, let us revisit the Hamiltonian
(\ref{eq:H_Jacobian}). If we assume that \emph{only two} of the three
bosons bind, say $\left|r_{1}\right|\ll1$ (up to permutation symmetry
$S_{+}$), then we end up with two decoupled Hamiltonians, \[
H_{\mathrm{rel}}\approx\!\left[\frac{1}{2}p_{r,1}^{2}+\frac{g}{\sqrt{2}}\delta(r_{1})\right]+\left[\frac{1}{2}p_{r,2}^{2}+\frac{1}{2}r_{2}^{2}+2\sqrt{\frac{2}{3}}g\delta(r_{2})\right]\!,\]
one for a free-space molecule ($r_{1}$, relating to the relative
\emph{ground} state described in Fig.~\ref{fig:2p_HO-states}a) and
one for an effective particle in an \emph{excited} state of an oscillator
with a delta-type dimple at the origin ($r_{2}$, the lowest excitation
corresponding to Fig.~\ref{fig:2p_HO-states}b). This yields the
relative wave function (excluding the trivial CM factor) \[
\psi(X)\propto S_{+}\left\{ e^{-|x_{1}-x_{2}|/a}U\left(-\epsilon,\frac{1}{2}(x_{1}+x_{2})-x_{3}\right)\right\} ,\]
where the parameters are determined in analogy to Eq.~(\ref{eq:pp-psi}).
We have checked that this expression qualitatively reproduces the
two-body pattern in Fig.~\ref{cap:2p-corr} for $g\to-\infty$. That
makes it tempting to think of the hybrid state as a hard-core {}``gas''
of a dimer---clumped near the trap center---and a third, unbound boson.
In that regime, the energy scales as $E(g)\sim\mathcal{N}+\frac{N}{2}-\frac{1}{2a^{2}}+\alpha$,
where $\lim_{g\to-\infty}\epsilon(g)-\frac{1}{2}\equiv\alpha\in2\mathbb{N}-1$.
In general, there can be different combinations of relative and CM
excitations ($\mathcal{N}\neq0$) which give nearly the same energy
-- this explains the splittings of all but the lowest hybrid levels
in Fig.~\ref{cap:energy}.

\subsection{Fermionizing states\label{sub:Fermionizing-states}}

Let us now focus on the highest level in Fig.~\ref{cap:energy},
which is the energetically lowest gas-like state. Its energy does
not diverge quadratically with $g\to-\infty$, but rather saturates.
By the fermionization map above, its limit is simply the energy of
$N$ free fermions, $\lim_{g\to-\infty}E(g)=\sum_{a=0}^{N-1}\epsilon_{a}=\frac{N^{2}}{2}$
(see Fig.~\ref{cap:energy}, inset), and likewise for higher excitations.
Evidently, this requires a huge energy for the connecting level $E(g=0)>E(-\infty)$
if $N\gg1$. In fact, the difference between the two equals that between
the noninteracting \emph{ground state} $N\epsilon_{0}$ and its fermionization
limit $\sum_{a=0}^{N-1}\epsilon_{a}$, which can be written down explicitly
in a harmonic trap, \[
\left|E(g\to-\infty)-E(0)\right|=\frac{N^{2}}{2}-\frac{N}{2}=\frac{N(N-1)}{2}.\]
This can be thought of as increasing (lowering) the energy by $\Delta\epsilon=1$
for each pair $(i<j)$. Therefore, the {}``super-Tonks'' level connects
to the noninteracting level $E(0)=\frac{N^{2}}{2}+\frac{N(N-1)}{2}=N(N-\frac{1}{2})$,
as may be verified for $N\le4$ in Fig.~\ref{cap:energy}(inset). 

Accordingly, the corresponding many-body state is expected to evolve
to (\ref{eq:TG-HO}). A glimpse of this can be gotten from the density
profile shown in Fig.~\ref{cap:density}(c), where the $N$ density
wiggles characteristic of the Tonks gas emerge, $\rho=\sum_{a=0}^{N-1}|\phi_{a}|^{2}$
(a similar observation has been stated in Ref.~\cite{astrakharchik04a}).
In the repulsive case, this has the familiar interpretation that the
$N$ bosons localize on $N$ more or less {}``discrete'' spots due
to a trade-off between mutual isolation and external confinement \cite{deuretzbacher06,zoellner06a}. 

Some more insight into this crossover is given by the two-body correlation
function $\rho_{2}(x_{1},x_{2})$ (Fig.~\ref{cap:2p-corr}). As expected
from the two-atom toy model (Sec.~\ref{sub:Illustration}), the diagonal
$\{ x_{1}=x_{2}\}$ {}``damps out'' more and more. The fact that
it persists even for couplings as large as $g=-10$ underscores the
notion of the (finite-$g$) super-Tonks gas being more strongly correlated
than its repulsive counterpart \cite{astrakharchik05}. This in turn
relates to the picture that, due to a positive 1D scattering length
$a=-2/g$, a small region is excluded from the scattering zone, so
that the hard core effectively extends to a nonzero volume. This also
offers an explanation for another phenomenon: As $g\to-\infty$, we
see that the typical fermionized checkerboard pattern forms in the
two-body density \cite{girardeau01,zoellner06b}. This signifies that,
upon measuring a first boson at, say, $x_{1}\approx0$, the remaining
$N-1=2$ bosons are pinpointed to discrete positions $x_{2}\approx\pm1.5$.
However, here the peaks are much more pronounced than in the Tonks
gas, which may be accounted for by a {}``thicker'' hard core between
the atoms.\\

\begin{figure*}[!]
\subfigure[]{\includegraphics[width=0.33\textwidth,keepaspectratio]{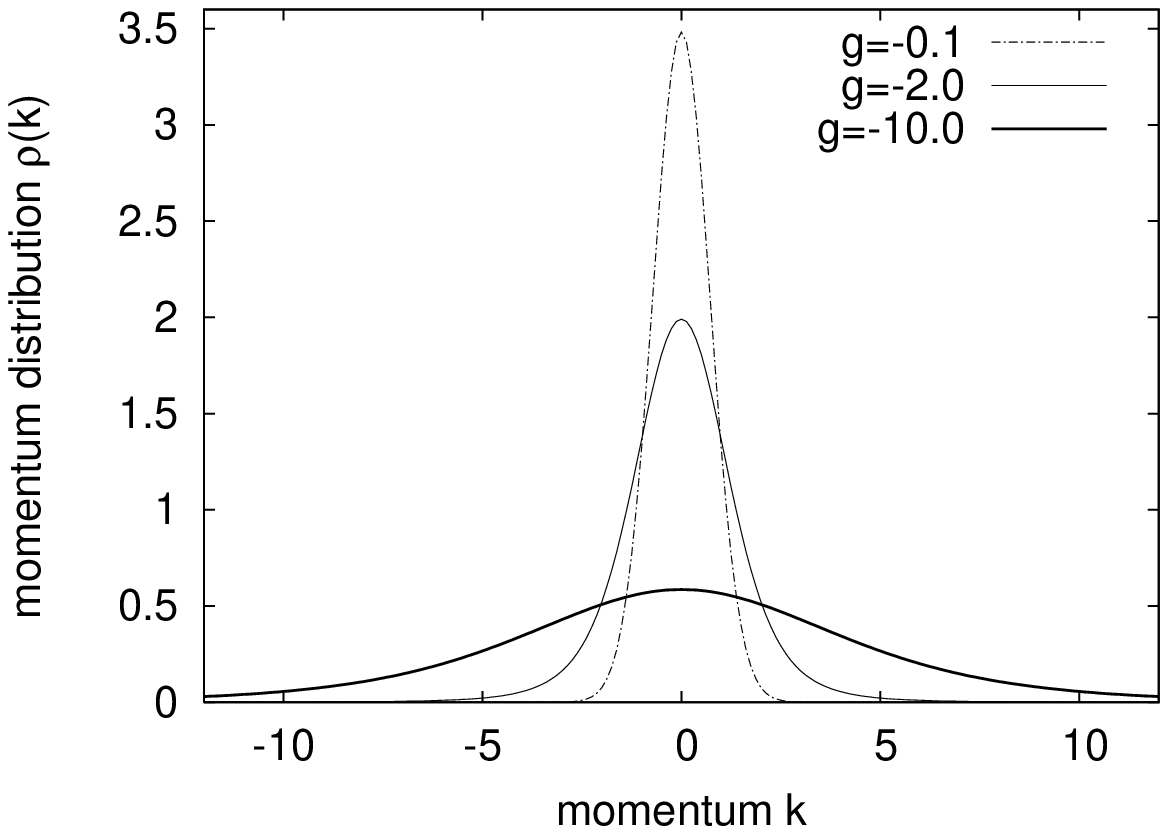}}\subfigure[]{\includegraphics[width=0.33\textwidth,keepaspectratio]{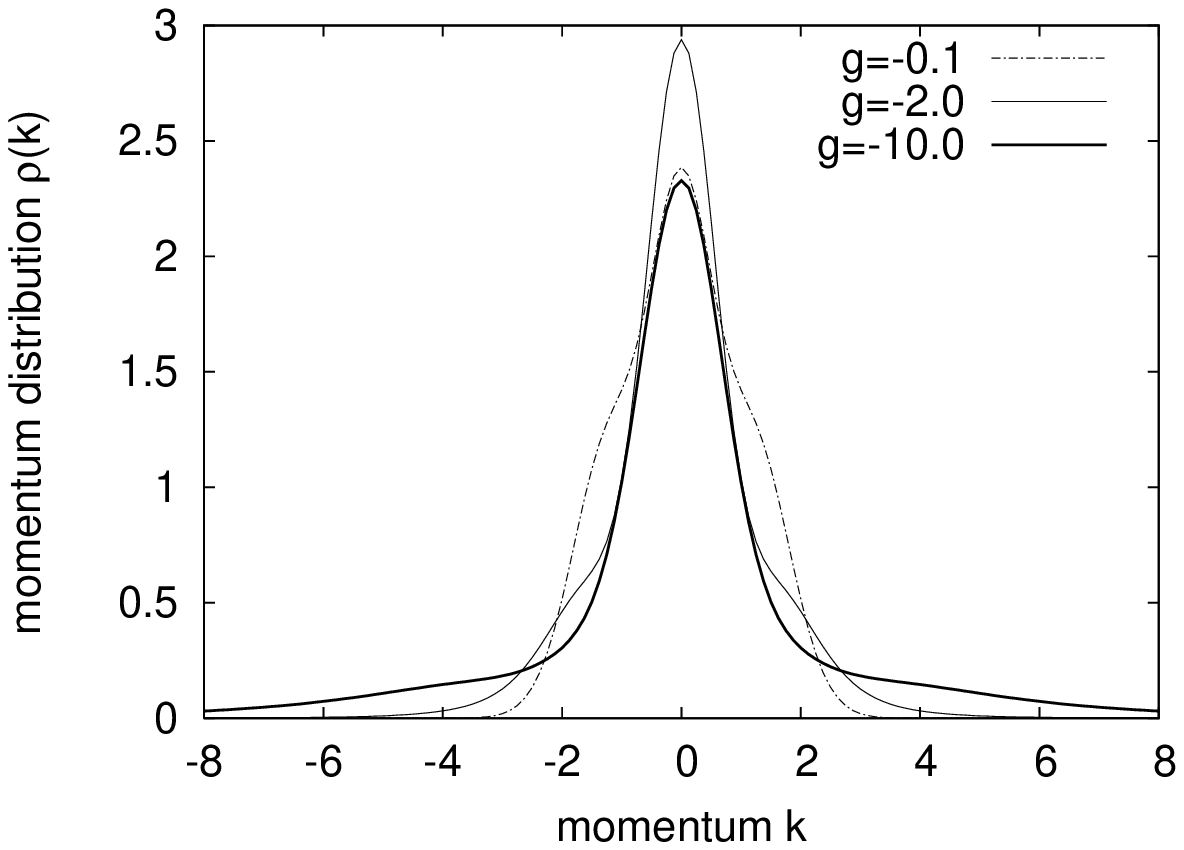}}\subfigure[]{\includegraphics[width=0.33\textwidth,keepaspectratio]{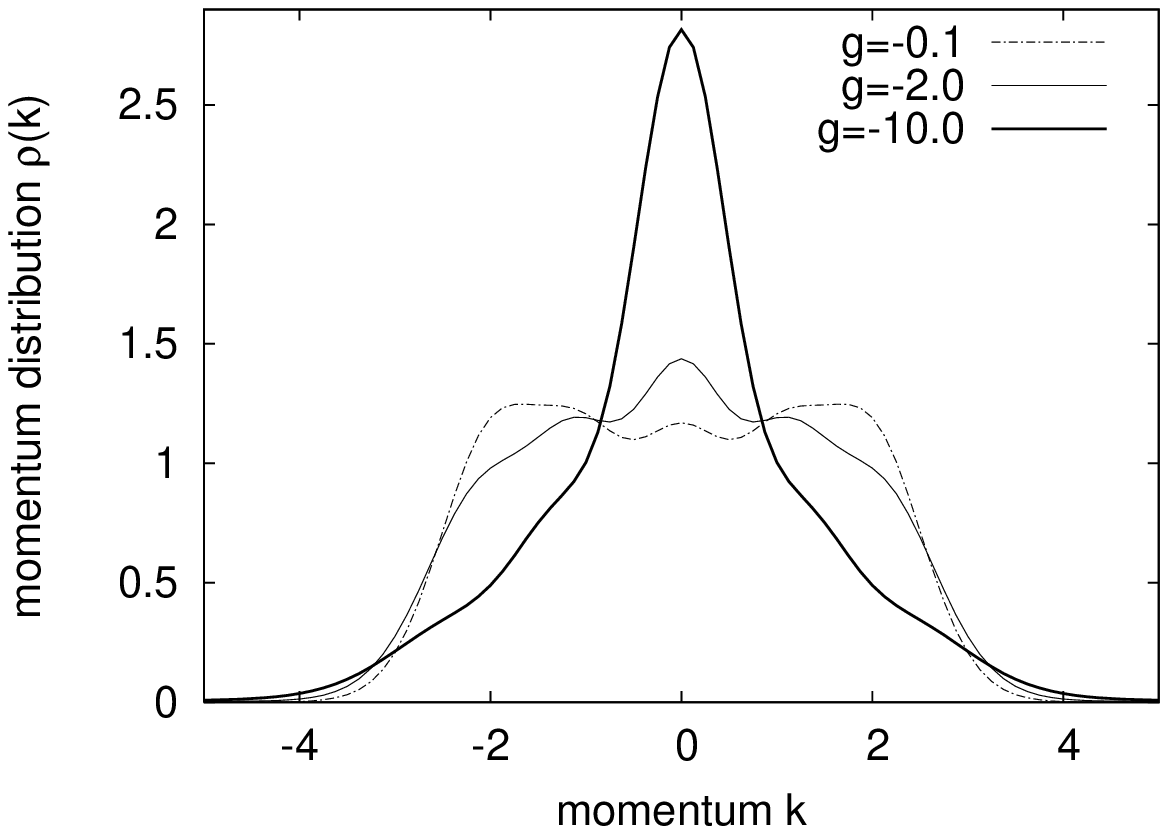}}

\caption{Momentum distribution $\tilde{\rho}(k)$ for the ground state (a),
the lowest hybrid state (b), and the lowest fermionizing state \textbf{(}c)
of $N=3$ bosons, plotted for different interaction strengths $g$.
\label{cap:momentum}}
\end{figure*}

\noindent We have so far looked into local observables, where the
fermionization limit imprints truly fermionic properties on the bosons.
By contrast, \emph{nonlocal} features as evidenced, e.g., in the experimentally
relevant momentum distribution\[
\tilde{\rho}(k)=2\pi\langle k|\rho_{1}|k\rangle=\int dx\int dx'e^{-ik(x-x')}\rho_{1}(x,x')\]
reflect nontrivial differences from the ideal Fermi gas. Figure~\ref{cap:momentum}(c)
shows the evolution of $\tilde{\rho}(k)$ for the gas-like state in
juxtaposition with the trimer and dimer states {[}Figs.~\ref{cap:momentum}(a-b)].
For the ground state, where all bosons simply form an ever tighter
$N$-body molecule as $g\to-\infty$, $\rho_{1}(x,x')\propto\delta(x-x')$
loses all long-range order, i.e., $\rho_{1}(x,x')=0$ for $|x-x'|>0$.
By complementarity, its momentum spectrum trivially approaches a flat
shape. Things are more complicated for the hybrid state in Fig.~\ref{cap:momentum}(b):
Since only two bosons bind and, as a whole, form a hard-core composite
with the remaining atom, some long-range order is preserved, so that
the central peak $\tilde{\rho}(0)$ persists even for large values
of $|g|$. Note that, as in the repulsive case, the hard-core short-range
correlations enforce an algebraic decay for high momenta, $\tilde{\rho}(k)\sim c/k^{4}$
\cite{minguzzi02}. Finally, the gas-like state exhibits the most
interesting behavior (Fig.~\ref{cap:momentum}c). The initially box-like
distribution $\tilde{\rho}(k)=2\pi\rho(k)$ (harmonic trap at $g=0$)
forms a strong $k=0$ peak highly reminiscent of the Tonks gas \cite{deuretzbacher06,zoellner06b}.
Also note the slow formation of the characteristic $k^{-4}$ tails.
This complements our picture of the crossover from zero to strongly
attractive interactions.

\section{Summary}

In this work, we have brought together the subjects of attractive,
one-dimensional Bose gases---which currently are of great interest
and experimentally relevant---and the binding properties of few-body
systems. We have studied the stationary states of one-dimensional
bosons in a harmonic trap throughout the crossover from weakly to
strongly attractive interactions. 

Three different classes of states have emerged for strong enough attraction:
(i) The ground state and its center-of-mass excitations become $N$-body
bound states, for which any two particles pair up to a tightly bound
molecule. Its binding length $a$ shrinks to zero with increasing
attraction, and thus the relative motion becomes independent of the
trap geometry. (ii) By contrast, certain highly excited states fermionize,
i.e., they map to an ideal Fermi state for infinite attraction. Both
the typical fermionic density profile with $N$ maxima and, more generally,
the characteristic checkerboard pattern in the two-body density have
been evidenced, signifying localization of the individual atoms. Also,
the formation of a hard-core momentum distribution has been witnessed,
with a zero-momentum peak and an algebraic decay for large momenta.
(iii) Between these two extremes, there is a rich class of hybrid
states featuring mixed molecule and hard-core boundary conditions.
For the special case of $N=3$ atoms, this class consists of a dimer
plus a single boson, with a hard-core separation between the two.

Even though we have focused mostly on few atoms ($N=3$) in a harmonic
trap, these results reflect the microscopic mechanism for arbitrary
atom numbers and external potentials. 

\begin{acknowledgments}
Financial support from the Landesstiftung Baden-Württemberg through
the project {}``Mesoscopics and atom optics of small ensembles of
ultracold atoms'' is gratefully acknowledged by PS and SZ. We thank
H.-D. Meyer, F. Deuretzbacher, and X. W. Guan \textbf{}for fruitful
discussions.
\end{acknowledgments}
\bibliographystyle{prsty}
\bibliography{/home/sascha/paper/pra/DW/phd,/home/sascha/bib/mctdh}

\end{document}